\documentstyle [12pt]{article}
\newcommand{\qqbar}{Q\overline{Q}}
\newcommand{\KKbar}{K\overline{K}}

\newcommand{\nnbar}{\overline{n}n}

\newcommand{\ssbar}{s\overline{s}}

\newcommand{\bea}{\begin{eqnarray}}
\newcommand{\eea}{\end{eqnarray}}
\newcommand{\be}{\begin{equation}}
\newcommand{\ee}{\end{equation}}

\newcommand{\gapproxeq}{\lower
.7ex\hbox{$\;\stackrel{\textstyle >}{\sim}\;$}}
\newcommand{\lapproxeq}{\lower
.7ex\hbox{$\;\stackrel{\textstyle <}{\sim}\;$}}

\begin{document}
\title{\small \rm \begin{flushright} \small{hep-ph/9511442}\\
\small{RAL-TR-95-070}\\
Date October 1995 \end{flushright} \vspace{2cm}
\LARGE {\bf From Coloured Quarks to Quarkonia, Glueballs and Hybrids}
\vspace{0.8cm} }
\author{ Frank E. Close\thanks{E-mail : fec@v2.rl.ac.uk} \\
{\small \em Particle Theory, Rutherford-Appleton Laboratory,
Chilton,
Didcot OX11 0QX, UK} \\  \\}
\date{October 1995 \vspace{1.0cm}}

\begin{center}
\maketitle

\begin{abstract}
Lectures at VII Jorge Andre Swieca Summer School in Nuclear Physics; Brasil;
Jan 22 - Feb 4 1995. Published by Plenum Press (C.A.Amaral Nunes ed.)
\end{abstract}
\end{center}

\newpage

\section{COMPOSITE SYSTEMS: FROM MOLECULES TO QUARKS}

The first sign that there is an underlying structure at some level
of matter is the existence of an excitation spectrum. Thus molecules
exhibit a spectrum due to the quantised motions of their constituent atoms.
In turn atomic spectra are due to their electrons, those of nuclei are due to
their constituent protons and neutrons and, we now realise, those of the
hadrons are due to their constituent quarks. Qualitatively these look similar
but quantitatively they are radically different. Molecular excitations are on
the scale of $meV$, atoms $eV$, nuclei $MeV$ (note $M$ versus $m$ !) and the
hadrons are hundreds of $MeV$. These represent, inter alia, the different
length scales and binding energies of the respective structures. They also
are related to the different momentum or energy scales needed for probes to
resolve substructures directly.

The discovery of the atomic nucleus was achieved with probes ($\alpha$
particles) that are provided by
natural radioactive sources. These resolved the atom and revealed its
nucleus but saw the latter only as  a point of charge; the inner structure
was not then apparent. If the beams are, say, electrons that have
been accelerated to energies of
some hundreds of $MeV$, the protons and neutrons become visible as
individuals. If the beams have energies of tens to hundreds of $GeV$, the
inner structure of the nucleons is resolved and their quarks are directly seen.

I do not wish here to enter into a debate on the detailed relation between
quarks as revealed in the latter ``deep inelastic" experiments and those
that drive spectroscop: it is the latter on which I shall concentrate.
In particular I shall motivate the exciting possibility that new varieties
 of hadron are emerging which are associated with the
degrees of freedom available
to the force fields that bind quarks.

So first we need to ask what force holds hadrons together.

The quarks carry electrical charges but also carry an extra charge called
``colour". There are three varieties, let's call them red blue green, and they
attract and repel as do electrical charges: like colours repel
and  unlike attract (technically when in
antisymmetric quantum states). Thus three different colours can mutually
attract
and form a baryon but a fourth is left neutral: attracted by two and repelled
 by the third. The restriction of attractions to the antisymmetric state
causes the attraction and repulsion to counterbalance, (the importance
 of symmetry will be discussed in section 2).

The analogy between colour charge and electrical charge goes further.

By analogy with QED one can form QCD, quantum chromo(or colour) dynamics.
Instead of photons as radiation and force carriers one has (coloured)
gluons. The gluons are in general coloured because a quark that is
coloured Red (say) can turn into a Green by radiating a gluon that is
coloured ``Red-Green"; (technically the $3$ colours form the basis
representation of an $SU(3)$ group and the gluons transform as the
regular representation, the octet - see section 2).
  The fact that the gluons carry the colour, or charge, whereas
photons do not carry any charge, causes gluons to propagate differently from
photons. Whereas photons can voyage independently, gluons can mutually
attract en route (they ``shine in their own light"). Not only does this affect
the long range behaviour of the forces but it also suggests that bound states
of pure glue, known as glueballs, may exist. The existence of glueballs and
other
hadrons where glue is excited (``hybrids") will be the focus of these lectures.
First I shall discuss the ways that colour manifests itself in more familiar
systems such as baryons and nuclei.

To illustrate the similarities and differences between QED and QCD we can
list particles and clusters according to whether they feel the force or not.
Those that feel the force may do so because they manifestly carry the charge
(such as electrons or ions in QED or quarks and gluons in QCD) or because
the charge is hidden internally (such as atoms and molecules in QED or
nucleons and nuclei in QCD). Contrast these systems
with those states that do not feel the
force directly as they neither carry nor
contain the charge (such as neutrinos and photons
in QED or leptons in QCD). Note in particular that the force carriers, the
photon and gluon, are in different parts of this matrix; the gluons have
manifest colour charge and feel the QCD forces whereas the photons do not
have electrcial charge and do not directly feel the QED forces.

We can go further and examine the particular set of systems with ``hidden"
charge within them. There are three broad classes of these ``consequential"
forces. In QED the atoms and molecules feel covalent, van der Waals and ionic
forces; the former pair being due essentially to constituent exchange
and two-photon exchange between two separate pairs of constituents
respectively.
The analogues in QCD are quark exchange and two-gluon exchange; there is no
analogue of ionic forces at long range due to the property of confinement
of colour in QCD.

The confinement also breaks the naive similarity between QED and QCD forces
in that the quark exchange (``covalent" force) involves clusters or bags
of colourless combinations of
quark and antiquark known as mesons (of which the light
pion is the most obvious in nuclear forces). Nonetheless we can imagine
nuclei as being the QCD analogues of molecules. The van der Waal, two gluon
exchange, forces
 will also be affected by confinement and, presumably, will involve
glueballs as effective exchange objects. However, as the lightest glueballs
are not expected to exist below 1GeV, their effective range is very restricted
and so they are unlikely to affect nuclear forces in any immediately
observable way. The search for colour analogues of van der Waals forces
is likely to be unfruitful in my opinion until someone comes up with a
smart idea.

I shall first show the important role that colour plays for quarks in baryons
and then I shall
contrast baryons, made of three constituent quarks, with the nuclei $^3H$
 and $^3He$ which are made of three nucleons. Then I shall concentrate on
colour in mesons, extracting information about the colour forces from
spectroscopy
(section 3). I shall then discuss decays (section 4) in order to contrast with
glueball decays (section 5) before considering a ``realistic" picture where
glueball and quarkonia mix (section 6). The production of glueballs is
discussed in section 7, the phenomenology of hybrids in section 8 and some
attempts to test the hybrid interpretation are in the final section.

\section{COLOUR, THE PAULI PRINCIPLE AND SPIN-FLAVOUR CORRELATIONS}
\subsection{Colour}

The Pauli principle forbids  fermions to coexist in the same
quantum state. Historically this created a paradox in that the baryon
$
\Omega^- (S^\uparrow S^\uparrow
S^\uparrow)$ appeared manifestly to violate this.

If quarks possess a property called colour, any quark being able to
carry any one of three colours (say red, green, blue), then the
$\Omega^-$ (and any baryon) can be built from distinguishable
quarks:
$$
\Omega^- (S^\uparrow_R S^\uparrow_G
S^\uparrow_B).
 $$
This was how the idea of threefold colour first entered particle physics.
Subsequently the idea developed that colour is the source of a relativistic
quantum field theory, QCD or quantum chromodynamics, and is the source of
the strong forces that bind quarks in hadrons. I shall first discuss the idea
of colour and how, when combined with the Pauli principle, it determines
the properties of baryons. Then I shall develop the idea of it as source
of the interquark forces.

If quarks carry colour but leptons do not, then it is natural to
speculate that colour may be the property that is the source of the
strong interquark forces - absent for leptons.

Electric charges obey the rule ``like repel, unlike attract" and cluster
to net uncharged systems.  Colours obey a similar rule: ``like
colours repel, unlike (can) attract".  If the three colours form the
basis of an SU(3) group, then they cluster to form ``white" systems -
viz. the singlets of SU(3).  Given a random soup of coloured
quarks, the attractions gather them into white clusters, at which
point the colour forces are saturated. Nuclear forces are then the
residual forces among these clusters.

If quark ($Q$) and antiquark ($\bar{Q}$) are the $\underline{3}$
and
$\underline{\bar{3}}$ of colour SU(3), then combining up to three
together gives SU(3) multiplets of dimensions as follows (see e.g.
ref.\cite{Closebook}):
\bea
QQ = \mathop{3}_{\sim} \times
\mathop{3}_{\sim} = \mathop{6}_{\sim}
+ \mathop{\bar{3}}_{\sim}\nonumber \\
Q\bar{Q} = \mathop{3}_{\sim}  \times \mathop{\bar{3}}_{\sim}
=
\mathop{8}_{\sim}+ \mathop{1}_{\sim}\nonumber
\eea
The $Q\bar{Q}$ contains a singlet - the physical mesons. Coloured
gluons belong to the $\mathop{8}$ representation and are confined.
  Combining $QQ$ with
a third $Q$ gives \relax
$$
QQQ = \mathop{10}_{\sim}+ \mathop{8}_{\sim}+
\mathop{8}_{\sim}+\mathop{1}_{\sim}.
$$
where the $\mathop{1}\limits_{\sim}$ arose when $QQ$ pairs were in
$\mathop{\bar{3}}\limits_{\sim}$.

Note the singlet in $QQQ$ - the physical baryons.

For clusters of three or less, only $Q\bar{Q}$ and $QQQ$ contain
colour singlets and, moreover, these are the only states realized
physically.  Thus are we led to hypothesize that only colour singlets
can exist free in the laboratory; in particular, the quarks will not
exist as free particles.

\subsection{Symmetries and correlations in baryons}

To have three quarks in colour singlet:
$$
1\equiv\frac{1}{\sqrt{6}} [(RB-BR)Y + (YR-RY)B + (BY-YB)R]
$$
any pair is in the $\mathop{\bar{3}}\limits_{\sim}$ and is
antisymmetric.  Note that $\mathop{3}\limits_{\sim} \times
\mathop{3}\limits_{\sim} = \mathop{6}\limits_{\sim}
+\mathop{\bar{3}}\limits_{\sim}$.  These are explicitly
\bea
{\mathop{\bar{3}}\limits_{\sim}}_{anti}&
{\mathop{6}\limits_{\sim}}_{sym}\nonumber \\
RB-BR & RB+BR\nonumber \\
RY-YR & RY+YR\nonumber \\
BY-YB & BY+YB\nonumber \\
& RR\nonumber \\
&BB\nonumber \\
& YY
\eea
Note well: \underline{Any Pair is Colour Antisymmetric}\\

The Pauli principle requires total antisymmetry and therefore any
pair must be:

\underline{Symmetric in all else} (``else" means ``apart from colour").\\

This is an important difference from nuclear clusters where the
nucleons have no colour (hence are trivially \underline
{symmetric} in colour!).  Hence for nucleons Pauli says
\be
\mbox{\underline{Nucleons are Antisymmetric in Pairs}}
\ee
and for quarks
\be
\mbox{\underline{Quarks are Symmetric in Pairs}}
\ee
(in all apart from colour).

If we forget about colour (colour has taken care of the
antisymmetry and won't affect us again), then
(i) Two quarks can couple their spins as follows
\be
\left\{\begin{array}{rc}S = 1: & \mbox{symmetric}\\
 S=0: &\mbox{antisymmetric}\end{array}\right\}
\ee
(ii) Two $u,d$ quarks similarly form isospin states
\be
\left\{\begin{array}{rc}I=1: &\mbox{symmetric} \\
 I=0: &\mbox{antisymmetric}\end{array}\right\}
\ee
(iii) In the ground state $L=0$ for all quarks; hence the orbital state
is trivially symmetric.  Thus for pairs in $L=0$, we have due to Pauli that
\be
\left\{\begin{array}{rcrc}S=1 & \mbox{and} & I=1 &
\mbox{correlate} \\
 S=0& \mbox{and} & I=0 & \mbox{correlate}\end{array}\right\}
\ee
Thus the $\Sigma^0$ and $\Lambda^0$ which are distinguished
by
their $u,d$ being $I=1$ or $0$ respectively also have the $u,d$ pair
in spin $=1$ or $0$ respectively:
\be
\left\{\begin{array}{rcr}\Sigma^0[S(u,d)_{I=1}] &\leftrightarrow
&
S(u,d)_{S=1} \\
\Lambda^0[S(u,d)_{I=0}] &\leftrightarrow &
S(u,d)_{S=0}\end{array}\right\}
\ee
Thus, the spin of the $\Lambda^0$ is carried entirely by the strange
quark.

\subsection{Colour, the Pauli principle and magnetic moments}

The electrical charge of a baryon is the sum of its constituent quark
charges.  The magnetic moment is an intimate probe of the
correlations between the charges and spins of the constituents.
Being wise, today we can say that the neutron magnetic moment
was the first clue that the nucleons are not elementary particles.
Conversely the facts that quarks appear to have $g\simeq 2$
suggests that they {\it are} elementary (or that new dynamics is at
work if composite).

A very beautiful demonstration of symmetry at work is the
magnetic moment of two similar sets of systems of three, viz.
\bea
\left\{\begin{array}{ccl}N ;& P \\
ddu ;& uud\end{array}\right\}
&\mu_p/\mu_N = -3/2\nonumber
\eea
and the nuclei
\bea
\left\{\begin{array}{ccl}H^3;& He^3\\
NNP;& PPN\end{array}\right\}
 &\mu_{He}/\mu_H = -2/3\nonumber
\eea
The Pauli principle for nucleons requires $He^4$ to have {\it no}
magnetic moment:
$$
\mu[He^4;P^\uparrow P^\downarrow N^\uparrow
N^\downarrow ] = 0.
$$
Then
\bea
He^3 &\equiv He^4 - N\nonumber \\
H^3 &\equiv He^4 - P\nonumber
\eea
and so
$$
\frac{\mu_{He^3}}{\mu_{H^3}} =\frac{\mu_N}{\mu_p}
$$

To get at this result in a way that will bring best comparison with
the nucleon three-quark example, let's study the $He^3$ directly.

$He^3 = ppn: pp$ are flavour symmetric; hence Pauli requires that they be spin
antisymmetric; i.e., $S=0$.

Thus
\be
[He^3]^\uparrow\equiv (pp)_0 n^\uparrow
\ee
and so the $pp$ do not contribute to its magnetic moment.  The
magnetic moment (up to mass scale factors) is
\be
\mu_{He^3} = 0+\mu_n.
\ee
Similarly,
\be
\mu_{H^3} = 0+\mu_p.
\ee
which, of course, gives the result that we got before, as it must. But
deriving it this way is instructive as we see when we
study the nucleons in an analogous manner.

The proton contains $u,u$ flavour symmetric and {\it colour
antisymmetric}; thus the spin of the ``like" pair is symmetric
$(S=1)$ in contrast to the nuclear example where this pair had
$S=0$.  Thus coupling spin $1$ and spin $1/2$ together, the
Clebsches yield (where subscripts denote $S_z$)
\be
p^\uparrow =\frac{1}{\sqrt{3}} (u,u)_0d^\uparrow
+\frac{2}{\sqrt{3}} (u,u)_1d^\downarrow
\ee
(contrast Eq. (8)), and (up to mass factors)
\be
\mu_p =\frac{1}{3} (0+d) +\frac{2}{3} (2u-d).
\ee
Suppose that $\mu_{u,d} \propto e_{u,d}$, then
\be
\mu_\mu = -2\mu_d
\ee
so
\be
\frac{\mu_p}{\mu_N} =\frac{4u-d}{4d-u} =-\frac{3}{2}
\ee
(the neutron follows from proton by replacing $u\leftrightarrow
d$).

I cannot overstress the crucial, hidden role that colour played here
in getting the flavour-spin correlation right.

\section{THE POTENTIAL AND THE FORCE}

The following remarks are by no means rigorous and are intended only to
abstract some general suggestive features about the dynamics from the
spectroscopy of hadrons. They will also enable us to draw up some
empirical guidelines for identifying the nature of light hadrons.

We all know what the spectrum of a Coulomb potential looks like, with
the energy gap between the first two levels already being well on the way
to ionisation energies. The spectrum of hadrons is not like this in that
the gap between 1S and 2S is similar to (though slightly greater than) that
between 2S and 3S, and so on to 3S and 4S etc. The P states are found
slightly above the midway between the corresponding S states. This is
similar to that of a linear  potential (which is near enough to a
harmonic that for many purposes the latter is often used for analytical
calculations). A comparison is shown in fig 1.

It is instructive to use the particle data tables\cite{PDG}
 and to place the $b\bar{b}$
states on this spectrum, noting the relative energy gaps between the 1S,2S,
3S and the 1P,2P states. Now do the same for $c\bar{c}$ but rescaled
downwards by 6360MeV (so the $\psi(3097)$ and $\Upsilon(9460)$ start the
1S states at the same place). It is remarkable that where corresponding
levels have been identified in the two spectroscopies, there is a rather
similar pattern both qualitatively and even quantitatively (see figs 2a,b).

We shall consider the implications of this for light hadrons later but first
we can abstract the message that the potential between heavy flavours is
linear to a good approximation. This immediately tells us about the
spatial dynamics of the force fields. Let me show you how.

In the case of a U(1) charge, as in electrostatics, the force fields
spread out in space symmetrically in all three dimensions. Thus the
intensity crossing a sphere at distance R dies as the surface area, hence
as $\frac{1}{R^2}$. The potential is the integral of this, hence proportional
to $\frac{1}{R}$, the Coulomb form. We see that the Coulomb potential is
``natural" in a 3-D world.

Contrast this with the empirical message from the $Q\bar{Q}$ spectroscopy,
where $V(R) \sim R$. Here the intensity $\sim \frac{dV}{dR} \sim constant$.
The intensity does not spread at all; it is indeed ``linear". From this
empirical observation we have the picture that the gauge fields, the
gluons, transmit the force as if in a tube of colour flux. This is also
substantiated by computer simulations of QCD (``lattice QCD"). There is
some limited transverse spread but to a first approximation one is
encouraged towards models where a linear flux tube drives the dynamics.

This is what we find for the long range nature of the potential, where
the gluons have mutually interacted while transmitting the force. At
short range one expects there will be a significant perturbation
arising from single gluons travelling between the quarks independently;
this will be akin to the more familiar case of QED where independent
photon exchange generates the $\frac{1}{R}$ behaviour discussed above.
Hence our intuition is that the full potential in QCD will have a structure
along the lines of
\begin{equation}
V(R) \sim \frac{\alpha_s}{R} + aR
\end{equation}
where $\alpha_s$ is the strong coupling strength in QCD and $a$ is a
constant with dimensions of energy per unit length; this is in effect
the tension in the flux tube and empirically is about 1GeV/fermi.
This potential, when plotted on graph paper, looks similar to a log(R)
at the distance scales of hadrons. It is for this reason that the
absolute energy gap between 1S and 2S say is nearly independent of
the constituent mass: the solutions to the Schrodinger equation for
a log potential show that the energy gaps are independent of mass
(for $\frac{1}{R}$ they grow $\sim M$ whereas for $R$ they fall as
$M^{-\frac{1}{3}}$ and the competition ``accidentally" cancels.)

As an aside we can illustrate this.

Consider the Schrodinger equation

$$
(\nabla^2 + 2m_1 R^N) \psi_1(R) = 2m_1 E_1 \psi_1 (R)
$$
and similar form for $m_2$ with $E_2$ and $\psi_2 (R)$.  Let $R\rightarrow
\lambda R$ where $\psi_2 (R) \equiv  \psi_1 (\lambda  R)$.  Thus we
compare
$$
(\nabla^2 + 2m_1 \lambda^{N+2}  R^N) \psi_1(\lambda R) = 2m_1 E_1
\lambda^2 \psi_1 (\lambda R)
$$
with
$$
(\nabla^2 + 2m_2   R^N) \psi_2( R) = 2m_2 E_2  \psi_2 ( R)
$$

Recognising that $\psi_2 (R) \equiv \psi_1 (\lambda R)$, on the L.H.S. we have
$$
\lambda^2 \equiv (\frac{m_2}{m_1})^{2/(N+2)}
$$
and on the R.H.S.
$$
\frac{E_2}{E_1} \equiv \frac{m_1}{m_2} \lambda^2
$$
Hence
$$
\frac{E_2}{E_1} =  (\frac{m_2}{m_1})^{-\frac{N}{N+2}}
$$
shows how the energy levels scale with constituent mass in a potential $R^N$.

We can make a further analogy between QED and QCD via the magnetic
perturbations on the ground states. In hydrogen the magnetic interaction
between electron and proton causes a hyperfine splitting between the
$^3S_1$ and $^1S_0$ levels. This is inversely proportional to the
constituent masses and proportional to the expectation of the wavefunction
at the origin and to $\langle \vec{S_1} \cdot \vec{S_2} \rangle$.
 For mesons one finds a similar splitting where
for $Q\bar{q}$ states the $^3S_1$ and $^1S_0$ levels are as follows

\vskip 0.1in

\begin{tabular}{l|lllll}
& K & D & $D_s$ & B & $B_s$ \\
$m(^3S_1)$ & 0.89 & 2.01 & 2.11 & 5.32 & 5.33\\
$m(^1S_0)$  & 0.49 & 1.87 & 1.97 & 5.27 & 5.28 \\
\end{tabular}
(the vector is raised by 1 unit and the pseudoscalar reduced
by three units relative to the unperturbed values; this follows from
$\langle 2\vec{S_1}\cdot \vec{S_2}\rangle \equiv
\langle(\vec{S_1}+\vec{S_2})^2
- 2\vec{S_i}^2\rangle \equiv S(S+1) -\frac{3}{2})$.
Qualitatively we see that the magnitude of the splitting is smaller as
one proceeds to heavier flavours, in line with the inverse mass property
of (chromo)magnetic interactions. Quantitavely the behaviour is interesting.
For a potential $V(R) \sim R^N$ the wavefunction at the origin behaves as
\begin{equation}
\psi(0)^2 \sim \mu_{ij}^{\frac{3}{2+N}}
\end{equation}
where $\mu$ is the reduced mass
\begin{equation}
\frac{1}{\mu_{ij}} = \frac{1}{m_i} + \frac{1}{m_j}
\end{equation}
Now if we assume that
\begin{equation}
\frac{(m_V + m_P)_{ij}}{2} \equiv  m_i + m_j
\end{equation}
and note that, in hyperfine splitting
\begin{equation}
m_V - m_P \sim \frac{\psi(0)^2}{m_im_j}
\end{equation}
where $i,j$ are constituent quarks comprising Vector or Pseudoscalar mesons
($q_i\bar{q_j}$), we can find the best value of N in the potential.
If one forms $m_V^2 - m_P^2$ you will see that this is flavour independent to
a remarkable accuracy; then following the above hints you will immediately
see that $N=1$ is preferred; the wavefunctions of the linear potential are
those that fit best in the perturbation expression.

The mean mass of the ground states is nearer to the vector (spin triplet)
than the pseudoscalar (spin singlet). If we look at the
mass gap between the  $^3S_1$ 1S and the $^3P_2$ 1P levels, we find
again a remarkable flavour independence, not just for the $b\bar{b}$ and
$c\bar{c}$ already mentioned but for the strange and nonstrange too.

\begin{tabular}{l|rrrrrr}
& $u\bar{d}$ & $u\bar{s}$  & $s\bar{s}$  &$ c\bar{u}$  & $c\bar{c}$
&$b\bar{b}$
\\
$m(^3P_2) $& 1320 & 1430 & 1525 & 2460 & 3550 & 9915 \\
$m(^3S_1)$ & 770 & 892 & 1020 & 2010 & 3100 & 9460\\
gap & 550 & 540 & 500 & 450 & 450 & 450\\
\end{tabular}

Thus although the splittings between $^3S_1$ and $^1S_0$ are strongly mass
dependent, as expected in QCD, the $S-P$ mass gaps are to good approximation
fairly similar across the flavours. Even though the light flavoured states are
above threshold for decays into hadrons, the memory of the underlying
potential remains and, at least empirically, we can produce an ouline
skeleton for the spectroscopic pattern anticipated for all flavours.
I illustrate this in fig 2. The absolute separations of 1S, 2S, 3S and those
of 1P,2P have been taken from the known heavy flavours and rescaled slightly
to make a best fit where the 1D,1F and even 1G are found by the high
spin states in each of these levels. A numerical solution of the
spectrum in a model where $Q\bar{Q}$ are connected by a linear flux-tube is
shown in fig.3; this is indeed very similar to the data and empirical spectrum
illustrated in fig 2c.

Unless certain $J^{PC}$ have strong energy shifts through coupling to
open channels, this should give a reliable guide to the energies of
light hadron multiplets.When we combine the $q\bar{q}$ spins to singlet or
triplet ($S=0,1$) and then combine in turn with the orbital angular
momentum we can construct a set of $^{2S+1}L_J$ states. We shall be interested
later in the possible discovery of a scalar glueball and so we shall also
need to be aware that scalar mesons can be formed in the quark model as $^3P_0$
states. From the figure we anticipate these to lie in the region around $1.2
(n\bar{n}) - 1.6 (s\bar{s})$GeV.

A list of the low lying quarkonium multiplets is given below

\begin{table}[htbp]
\begin{center}
\begin{tabular}{l|ll}
& S=1 triplet & S=0 singlet \\\hline
S & $^3S_1 \; 1^{--}$ &   $^1S_0 \; 0^{-+} $\\\hline
P & $^3P_J \; 0^{++} 1^{++} 2^{++}$ &  $^1P_1 \; 1^{+-} $\\
\hline
D & $^3D_J \; 1^{--} 2^{--} 3^{--}$  &  $ ^1D_2 \; 2^{-+} $\\
\hline
F & $^3F_J \; 2^{++} 3^{++} 4^{++} $ &  $ ^1F_3 \; 3^{+-}$\\
\end{tabular}
\end{center}
\end{table}

The spectroscopy of
baryons and mesons is now rather well understood, at least in outline,
to an extent that if there are ``strangers" lurking among the conventional
states, there is a strong likelihood that they can be smoked out.

Such a hope is now becoming important as new states are appearing
and may have  a radical implication for our understanding of strong-QCD.
The reason has to do with the nature of gluons. As gluons carry colour charge
and can mutually attract, it is theoretically plausible that gluons can
form clusters that are overall colourless (like conventional hadrons)
but which contain only gluons. These are known as ``glueballs" and would
represent a new form of matter on the 1fm scale.

Glueballs are a missing link of the standard model.
Whereas the gluon degrees of freedom expressed
in $L_{QCD}$ have been established beyond doubt in high
momentum data, their dynamics in the strongly interacting limit
epitomised by hadron spectroscopy are quite obscure. This may be
about to change as a family of candidates for gluonic hadrons (glueballs and
hybrids) is now emerging \cite{amsler94,cafe95,cp94}. These contain both
hybrids around 1.9GeV and a scalar glueball candidate at $f_0(1500)$.

In advance of the most recent data, theoretical arguments
suggested that there may be gluonic activity manifested in the 1.5
GeV mass region. Lattice QCD is the best simulation of theory and
predicts the lightest ``primitive" (ie quenched approximation)
glueball to be $0^{++}$ with mass $1.55 \pm 0.05$ GeV
\cite{ukqcd}. Recent lattice computations place the glueball
slightly higher in mass at $1.74 \pm 0.07$ GeV \cite{weing}
with an optimised value for phenomenology proposed by Teper\cite{teper}
of $1.57 \pm 0.09$ GeV. That lattice QCD computations of the scalar glueball
mass are now concerned with such fine details represents considerable
advance in this field. Whatever the final concensus may be, these
results suggest that scalar mesons in
the 1.5 GeV region merit special attention.
Complementing this has been the growing realisation that there
are now too many $0^{++}$ mesons confirmed for them all to be
$\qqbar$ states \cite{PDG,amsler94,cafe95,close92}.

I will introduce some of my own prejudices about glueballs and how to
find them. I caution that we have no clear guide and so others may
have different suggestions. At this stage any of us, or none of us,
could be right. We have to do the best we can guided by experience.
It is indeed ironical that the lattice predicts that the lightest glueball
exists in the same region of mass as quarkonium states of the same
$J^{PC} =0^{++}$. If this is indeed the case in nature,
the phenomenology of glueballs may well be more subtle than
naive expectations currently predict.

We shall be interested later in the possible discovery of ``hybrid" states,
where
the gluonic fields are dynamically excited in presence of quarks. Among these
we shall be particularly interested in $0^{-+}, 1^{-+}, 2^{-+}, 1^{--}$ and
possibly $1^{++}$. Note that the $1^{-+}$ configuration does not occur for
$Q\bar{Q}$ and so discovery of such a resonant state would be direct evidence
for dynamics beyond the simple quark model. The other quantum numbers can
be
shared by hybrids and ordinary states. The mass of these lightest hybrids
is predicted to be around $1.9$GeV in a dynamical model where quarks are
connected by a flux tube. The numerical solution of the dynamics is
discussed in ref\cite{bcs} and endorses the earlier estimates by Isgur
and Paton\cite{paton85}.In fig.3 we see a comparison of the predicted
hybrid spectroscopy and that of the conventional states. The mass of the
$2^{-+}$ hybrid is predicted to be tantalisingly close to that of the
conventional
$^1D_2$ with which it shares the same overall $J^{PC}$ quantum numbers.
Comparison with fig.2 shows that this mass region is also near to that of
$3S$ states which include $0^{-+}$ and $1^{--}$, quantum numbers shared with
the lightest hybrids.  Furthermore, the $1^{++}$ hybrid shares quantum numbers
with the $2P (^3P_1)$ quarkonium and, following fig.2, we may anticipate that
here too a similarlity in mass may ensue. Thus on mass grounds alone it may
be hard to disentangle hybrids and glueballs from conventional states. It will
be important to investigate both the production and decay patterns of these
various objects.

As regards the decays, we need to study both the flavour dependence in a
multiplet and also the spin and other intrinsic dynamical dependences
that may help to distinguish conventional quarkonia from states where
the gluonic degrees of freedom are excited. We shall therefore first look at
the flavour dependence.

\section{QUARKONIUM DECAY AMPLITUDES}
Let's review some basics of the flavour dependence of
two body decays for a $q\bar{q}$ state of arbitrary flavour.
This will be helpful in assigning meosns to nonets and will also
help us to understand some general features of glueball decays.

Consider a quarkonium state
\begin{equation}
|\qqbar\rangle = {\rm cos}\alpha |n\bar{n}\rangle -
{\rm sin}\alpha |s\bar{s}\rangle
\end{equation}
where
\begin{equation}
n\bar{n} \equiv (u\bar{u} + d\bar{d})/\sqrt{2}.
\end{equation}
The mixing angle $\alpha$ is related to the usual nonet mixing
angle $\theta$ \cite{PDG} by the relation
\begin{equation}
\alpha = 54.7^{\circ} + \theta.
\end{equation}
For $\theta=0$ the quarkonium state becomes pure SU(3)$_f$
octet, while for $\theta=\pm 90^{\circ}$ it becomes pure singlet.
Ideal mixing occurs for $\theta=35.3^{\circ}$ (-54.7$^{\circ}$)
for which the quarkonium state becomes pure $\ssbar$
($\nnbar$).

In general we define
\begin{equation}
\eta = {\rm cos}\phi |n\bar{n}\rangle - {\rm sin}\phi
|s\bar{s}\rangle
\end{equation}
and
\begin{equation}
\eta' = {\rm sin} \phi |n\bar{n}\rangle +
{\rm cos} \phi |s\bar{s}\rangle
\end{equation}
with $\phi = 54.7^{\circ} +\theta_{PS}$, where $\theta_{PS}$ is the
usual octet-singlet mixing angle in SU(3)$_f$ basis where
\begin{equation}
\eta = {\rm cos}(\theta_{PS}) |\eta_8\rangle -
{\rm sin}(\theta_{PS}) |\eta_1\rangle,
\end{equation}
\begin{equation}
\eta' = {\rm sin}(\theta_{PS}) |\eta_8\rangle +
{\rm cos}(\theta_{PS}) |\eta_1\rangle.
\end{equation}

The decay of quarkonium into a pair of mesons
$ \qqbar \rightarrow M(Q\bar{q_i}) M(q_i\bar{Q})$
involves the creation of $q_i\bar{q_i}$ from the vacuum.
If the ratio of the matrix elements for the creation
of $s\bar{s}$ versus $u\bar{u}$ or $d\bar{d}$ is denoted by

\begin{equation}
\rho \equiv \frac{\langle 0|V|s\bar{s}\rangle}{ \langle
0|V|d\bar{d}\rangle},
\end{equation}
then the decay amplitudes of an isoscalar $0^{++}$ (or $2^{++}$)
are proportional to

\begin{eqnarray}
\langle\qqbar|V|\pi\pi\rangle & = & {\rm cos} \alpha
\nonumber\\
\langle\qqbar|V|\KKbar\rangle & = & {\rm cos}
\alpha (\rho - \sqrt{2} {\rm tan}
\alpha)/2 \nonumber\\
\langle\qqbar|V|\eta\eta\rangle & = & {\rm cos}
\alpha (1 - \rho \sqrt{2} {\rm tan}
\alpha )/2 \nonumber\\
\langle\qqbar|V|\eta\eta^\prime\rangle & = & {\rm cos}
\alpha (1 + \rho \sqrt{2} {\rm
tan} \alpha )/2. \nonumber\\
\label{eq:quarkonium0}
\end{eqnarray}

The corresponding decay amplitudes of the isovector are

\begin{eqnarray}
\langle\qqbar|V|\KKbar\rangle & = & \rho/2 \nonumber\\
\langle\qqbar|V|\pi\eta\rangle & = & 1/\sqrt{2} \nonumber\\
\langle\qqbar|V|\pi\eta^\prime\rangle & = & 1/\sqrt{2},
\nonumber\\
\label{eq:quarkonium1}
\end{eqnarray}
and those for $K^*$ decay

\begin{eqnarray}
\langle\qqbar|V|K\pi\rangle & = & \sqrt{3}/2 \nonumber\\
\langle\qqbar|V|K\eta\rangle & =
& (\sqrt{2}\rho - 1)/\sqrt{8} \nonumber\\
\langle\qqbar|V|K\eta^\prime\rangle & = & (\sqrt{2}\rho + 1)
/\sqrt{8}. \nonumber\\
\label{eq:quarkonium2}
\end{eqnarray}

For clarity of presentation we have presented
eqn. \ref{eq:quarkonium0},\ref{eq:quarkonium1} and
\ref{eq:quarkonium2} in the approximation where
$\eta \equiv (n\bar{n} - s\bar{s}) / \sqrt{2}$ and
$\eta' \equiv (n\bar{n} +  s\bar{s}) / \sqrt{2}$, i.e. for a
pseudoscalar mixing angle $\theta_{PS} \sim -10^{\circ}$ ($\phi =
45^{\circ}$). This is a useful mnemonic;  the full expressions for
arbitrary $\eta, \eta'$ mixing angles $\theta_{PS}$ are given in
ref.\cite{cafe95}. Exact SU(3)$_f$ flavour symmetry corresponds to $\rho =
1$; empirically $\rho \geq 0.8$ for well established nonets such as
$1^{--}$ and $2^{++}$ \cite{dok95,Godfrey}.

The partial width into a particular meson pair $M_iM_j$ may be
written as
\begin{equation}
\Gamma_{ij} = c_{ij}|M_{ij}|^2\times |F_{ij} (\vec{q})|^2 \times p.s.
(\vec{q})
\equiv \gamma^2_{ij} \times |F_{ij} (\vec{q})|^2 \times p.s.
(\vec{q})
\label{Gamma}
\end{equation}
where $p.s.(\vec{q})$ denotes the phase-space, $F_{ij} (\vec{q})$
are model-dependent form factors which are discussed in detail in
ref.\cite{cafe95}, $M_{ij}$ is the relevant
amplitude (eqn. \ref{eq:quarkonium0},\ref{eq:quarkonium1} or
\ref{eq:quarkonium2}) and $c_{ij}$ is a weighting factor arising
from the sum over the various charge combinations, namely 4 for
$K\bar{K}$, 3 for $\pi\pi$, 2 for $\eta\eta^\prime$ and 1 for
$\eta\eta$ for isoscalar decay (eqn. \ref{eq:quarkonium0}), 4 for
$K\bar{K}$, 2 for $\pi\eta$ and 2 for $\pi\eta^\prime$ for
isovector decay (eqn. \ref{eq:quarkonium1}) and 2 for $K^*$
decays (eqn. \ref{eq:quarkonium2}). The dependence of
$\gamma^2_{ij} = c_{ij}|M_{ij}|^2$ upon the mixing angle $\alpha$
is shown in fig. \ref{alpha}a  for the isoscalar decay in the case of
SU(3)$_f$ symmetry, $\rho = 1$.

The figure illustrates some general points.

An $s\bar{s}$ state corresponds to $\alpha = 90^0$ for which $\pi \pi$
vanishes. The $K\bar{K}$ vanishes when there is destructive interference
between $s\bar{s}$ and $n\bar{n}$; notice that the $\eta\eta$ tends to
vanish here also as it tends to be roughly $\frac{1}{4}$ of the $K\bar{K}$
independent of the mixing angle $\alpha$ (this would be an exact relation
for the ideal $\eta$ used in the text; the figure shows the results for
realistic $\eta$ flavour composition). This correlation between $\eta\eta$ and
$K\bar{K}$ is expected for any quarkonium state and a violation in data
will therefore be significant in helping identify ``strangers".

This pattern of decays is expected to hold true for any meson that
contains $q\bar{q}$ in its initial configuration. Thus it applies to
conventional or to hybrid multiplets and distinguishing between them will
depend on dynamical features associated with the gluonic excitation
or the spin states of the quarks. The case of glueballs is qualitatively
different in that there is no intrinsic flavour present initially and
so the pattern of decays will depend, inter alia, on the dynamics of
flavour creation.

The traditional assumption has been that as glueballs are flavour singlets,
their decays should be analogous to those of a flavour singlet quarkonium.
The case of a flavour singlet corresponds to $\alpha = -30^0$ (or $150^0$).
Here we see that $\eta \eta' \rightarrow 0$ and the other channels are
populated in proportion to their charge weighting (namely 4:3:1 for $K\bar{K}
:\pi\pi:\eta\eta$). A flavour singlet glueball would be expected to
show these ratios too if it decays through a flavour singlet
 intermediate state.

We can now look into the decays of glueballs by
finding examples of decays where gluons are already believed to play a role.
The data are sparse and do show consistency with the flavour singlet idea;
however, one must exercise caution before applying this too widely. I shall
first illustrate the flavour singlet phenomenon as it manifests itself
for gluonic systems at energies far from the mass scales of light-flavoured
quarkonium. Then I shall investigate what modifications may be expected
for glueballs at mass scales of 1-2GeV where quarkonium states with the same
$J^{PC}$ may contaminate the picture.

\section{PRIMITIVE GLUEBALL DECAYS}

The decays of $c\bar{c}$, in particular $\chi_{0,2}$, provide a
direct window on $G$ dynamics in the $0^{++},2^{++}$ channels
insofar as the hadronic decays are triggered by $c\bar{c}
\rightarrow gg \rightarrow Q\bar{Q}Q\bar{Q}$ (fig.
\ref{3graphs}a). It is necessary to keep in mind that these are in a
different kinematic region to that appropriate to our main
analysis but, nonetheless, they offer some insights into the gluon
dynamics. Mixing between hard gluons
and $0^{++}$, $2^{++}$ $Q\bar{Q}$ states (fig. \ref{3graphs}c)
is improbable at these energies as the latter 1 - 1.5 GeV
states will be far off their mass-shell. Furthermore,the
 narrow widths of $\chi_{0,2}$ are consistent with
the hypothesis that the 3.5 GeV region is remote from the
prominent
$0^+,2^+$ glueballs, $G$. Thus we
expect that the dominant decay dynamics is triggered by hard
gluons directly fragmenting into two independent $Q\bar{Q}$
pairs (fig. \ref{3graphs}a) or showering into lower energy gluons
(fig. \ref{3graphs}b). We consider the former case now; mixing
with $Q\bar{Q}$ (fig. 6c) and $G \rightarrow GG$ (fig. 6b) will be discussed in
section 6.

\subsection*{$ G \rightarrow QQ\bar{Q}\bar{Q}$}
This was discussed in ref. \cite{closerev} and
the relative amplitudes for the process shown in fig.
\ref{3graphs}a
read
\begin{eqnarray}
\langle G|V|\pi\pi\rangle & = & 1 \nonumber\\
\langle G|V|\KKbar\rangle & = &  R \nonumber\\
\langle G|V|\eta\eta\rangle & = & (1+  R^2)/2 \nonumber\\
\langle G|V|\eta\eta^\prime\rangle & = & (1- R^2)/2,
\nonumber\\
\label{b}
\end{eqnarray}
with generalizations for arbitrary pseudoscalar mixing angles
given in  ref.\cite{cafe95} and where $R \equiv \langle
g|V|s\bar{s}\rangle/\langle
g|V|d\bar{d}\rangle$. SU(3)$_f$ symmetry corresponds to
$R^2=1$. In this case the relative branching ratios (after weighting
by the number of charge combinations) for the decays
$\chi_{0,2} \rightarrow \pi\pi,\eta\eta,\eta\eta',K\bar{K}$
would be in the relative ratios 3 : 1 : 0 : 4. Data for $\chi_0$ are in
accord with this where the branching ratios are (in parts per mil)
\cite{PDG}:

\begin{eqnarray}
B(\pi^0\pi^0) & = & 3.1 \pm 0.6 \nonumber\\
\frac{1}{2}B(\pi^+\pi^-) & = & 3.7 \pm 1.1 \nonumber\\
\frac{1}{2}B(K^+K^-) & = & 3.5 \pm 1.2  \nonumber\\
B(\eta\eta) & = & 2.5 \pm 1.1. \nonumber\\
\label{chidata}
\end{eqnarray}

No signal has been reported for $\eta\eta'$.
Flavour symmetry is manifested in the decays of $\chi_2$ also:
\begin{eqnarray}
B(\pi^0\pi^0 )& = & 1.1 \pm 0.3 \nonumber\\
\frac{1}{2}B(\pi^+\pi^-) & = & 0.95 \pm 0.50 \nonumber\\
\frac{1}{2}B(K^+K^-) & = & 0.75 \pm 0.55  \nonumber\\
B(\eta\eta) & = & 0.8 \pm 0.5, \nonumber\\
\label{chidata2}
\end{eqnarray}
again in parts per mil. The channel $\eta\eta'$
has not been observed either.  These results are natural as they
involve hard gluons away from the kinematic region where $G$
bound states dominate the dynamics. If glueballs occur at lower
energies and mix with nearby $Q\bar{Q}$ states, this will in
general lead to a distortion of the branching ratios from the
``ideal" equal weighting values above. (A pedagogical example
will be igven in the next section). It will
also  cause significant mixing
between $n\bar{n}$ and $s\bar{s}$ in the quarkonium
eigenstates. Conversely, ``ideal" nonets, where the
quarkonium eigenstates are $n\bar{n}$ and $s\bar{s}$, are
expected to signal those $J^{PC}$ channels where the masses of the
prominent glueballs are remote from those of the quarkonia.

An example of this is the $2^{++}$ sector where the quarkonium
members are ``ideal" which suggests that $G$ mixing is nugatory in this
channel.
These data collectively suggest
that prominent $2^{++}$ glueballs are not in the $1.2 -
1.6$ GeV region which in turn is consistent with lattice calculations where
the mass of the $2^{++}$ primitive glueball is predicted to be larger than 2
GeV. The sighting of a $2^{++}$ state in the glueball favoured central
production, decaying into $\eta\eta$ with no significant $\pi\pi$\cite{2170}
could be the first evidence for this state. There are also interesting
signals from BES on a narrow state in this mass region seen in $\psi
\rightarrow \gamma MM$ where $MM$ refer to mesons pairs, $\pi\pi,
K\bar{K}$
with branching ratios consistent with flavour symmetry\cite{beijing}.

\section{$Q\bar{Q}$ AND GLUEBALL DECAYS IN STRONG COUPLING QCD}
In the strong coupling ($g\rightarrow\infty$) lattice formulation
of QCD, hadrons consist of quarks and flux links, or flux tubes, on
the lattice. ``Primitive" $Q\bar{Q}$ mesons consist of
a quark and antiquark connected by a tube of coloured
flux whereas primitive glueballs consist of a loop of flux
(fig. \ref{Rivsx}a,b) \cite{paton85}. For finite $g$ these eigenstates remain a
complete basis set for QCD but are perturbed by two types of
interaction \cite{kokoski87}:

\begin{enumerate}
\item
$V_1$ which creates a $Q$ and a $\bar{Q}$ at neighbouring lattice
sites, together with an elementary flux-tube connecting them, as
illustrated in fig. \ref{Rivsx}c,
\item
$V_2$ which creates or destroys a unit of flux around any
plaquette (where a plaquette is an elementary
square with links on its edges), illustrated in fig. \ref{Rivsx}d.
\end{enumerate}

The perturbation $V_1$ in leading order causes decays of
$Q\bar{Q}$ (fig. \ref{Rivsx}e) and also induces mixing
between  the ``primitive" glueball $(G_0)$ and  $Q\bar{Q}$
(fig. \ref{Rivsx}f). It is perturbation $V_2$ in leading order that
causes glueball decays and leads to a final state consisting of
$G_0G_0$ (fig. \ref{Rivsx}g); decays into $Q\bar{Q}$ pairs occur at
higher order, by application of the perturbation $V_1$ twice.  This
latter sequence effectively causes $G_0$ mixing with $Q\bar{Q}$
followed by its decay. Application of $V_1^2$  leads
to a $Q^2\bar{Q}^2$ intermediate state which then turns into
colour singlet mesons by quark rearrangement (fig.
\ref{3graphs}a); application of $V_2$ would lead to direct
coupling to glue in $\eta, \eta'$ or $V_2\times V_1^2$ to their
$\qqbar$ content (fig. \ref{3graphs}b).

The absolute magnitudes of these various contributions require
commitment to a detailed dynamics and are beyond the scope
of this first survey. We concentrate here on their {\bf relative}
contributions to the  various two body pseudoscalar meson final
states available to $0^{++}$ meson decays.For $Q\bar{Q}
\rightarrow Q\bar{q}q\bar{Q}$ decays induced by $V_1$, the
relative branching ratios are given in eqn. \ref{eq:quarkonium0}
where one identifies

\begin{equation}
\rho \equiv \frac{\langle Q\bar{s} s\bar{Q} |V_1| Q\bar{Q}
\rangle }{\langle Q\bar{d} d\bar{Q}|V_1| Q\bar{Q}\rangle}.
\label{rho}
\end{equation}
The magnitude of $\rho$ and its dependence on $J^{PC}$ is a
challenge for the lattice. We turn now to consider the effect of
$V_1$ on the initial ``primitive" glueball $G_0$. Here too we allow
for possible flavour dependence and define

\begin{equation}
R^2 \equiv  \frac{\langle s\bar{s} |V_1| G_0 \rangle}
{\langle d\bar{d} |V_1| G_0 \rangle }.
\label{r2}
\end{equation}

The lattice may eventually guide us on this magnitude and also on
the ratio $R^2/\rho$. In the absence of this information we shall
leave $R$ as free parameter and set $\rho=1$.

\subsection{Glueball-$\qqbar$ mixing at $O(V_1)$}
In this first orientation we shall consider mixing between $G_0$
(the primitive glueball state) and the quarkonia, $n\bar{n}$ and
$s\bar{s}$, at leading order in $V_1$ but will ignore that between
the two different quarkonia which is assumed to be higher order
perturbation.

The mixed glueball state is then

\begin{equation}
G = |G_0\rangle + \frac{|n\bar{n}\rangle \langle
n\bar{n}|V_1|G_0 \rangle}{E_{G_0}-E_{n\bar{n}}}
+ \frac{|s\bar{s}\rangle \langle s\bar{s}|V_1|G_0 \rangle}{E_{G_0}-
E_{s\bar{s}}}
\label{perturb}
\end{equation}
which may be written as

\begin{equation}
G = |G_0\rangle + \frac{\langle
n\bar{n}|V_1|G_0\rangle}{\sqrt{2}(E_{G_0}-
E_{n\bar{n}})}\{\sqrt{2}
|n\bar{n}\rangle +\omega R^2 |s\bar{s}\rangle  \}
\end{equation}
where

\begin{equation}
\omega \equiv \frac{E_{G_0}-E_{n\bar{n}}}{ E_{G_0}-E_{s\bar{s}}}
\label{omega}
\end{equation}
 is the ratio
of the energy denominators for the $n\bar{n}$ and $s\bar{s}$
intermediate states in old fashioned perturbation theory (fig.
\ref{3graphs}d).

Denoting the dimensionless mixing parameter by

\begin{equation}
\xi \equiv \frac{\langle d\bar{d}|V_1|G_0 \rangle }{E_{G_0}-
E_{n\bar{n}}},
\end{equation}
the eigenstate becomes, to leading order in the perturbation,

\begin{eqnarray}
N_G |G \rangle = |G_0 \rangle + \xi \{\sqrt{2}|n\bar{n}\rangle  +
 \omega R^2 |s\bar{s} \rangle \}
\equiv |G_0 \rangle + \sqrt{2}\xi |\qqbar\rangle \nonumber\\
\label{3states}
\end{eqnarray}
 with the normalization
\begin{eqnarray}
N_G  = \sqrt{1 + \xi^2 (2 + \omega^2 R^4)} ,\nonumber\\
\end{eqnarray}

Recalling our definition of quarkonium mixing

\begin{equation}
|\qqbar \rangle  = {\rm cos} \alpha |n\bar{n} \rangle -
{\rm sin} \alpha |s\bar{s}
\rangle
\end{equation}
we see that $G_0$ has mixed with an effective quarkonium of
mixing angle $\alpha$ where $\sqrt{2}{\rm tan} \alpha = -
\omega R^2$ (eqn . \ref{eq:quarkonium0}). For example, if
$\omega R^2 \equiv 1$, the SU(3)$_f$ flavour symmetry
maps a glueball onto quarkonium where tan$\alpha = -
1/\sqrt{2}$ hence $\theta=-90^{\circ}$, leading to the familiar
flavour singlet
\begin{equation}
|\qqbar \rangle = |u\bar{u} +d\bar{d} +s\bar{s} \rangle /\sqrt{3}.
\end{equation}

When the glueball is far removed in mass from the $Q\bar{Q}$,
$\omega \rightarrow 1$ and flavour symmetry ensues;
the $\chi_{0,2}$ decay and the $2^{++}$ analysis earlier
are examples of this ``ideal" situation. However, when $\omega
\neq 1$, as will tend to be the case when $G_0$ is in the vicinity
of the primitive $Q\bar{Q}$ nonet (the $0^{++}$ case of interest
here), significant distortion from naive flavour singlet can arise.

In particular lattice QCD
suggests that the ``primitive" scalar glueball $G_0$
lies at or above 1500 MeV, hence above the $I=1$ $Q\bar{Q}$
state $a_0(1450)$ and the (presumed) associated $n\bar{n}$
$f_0(1370)$. Hence $E_{G_0}-E_{n\bar{n}} >0$ in the numerator of
$\omega$. The $\Delta m = m_{s\bar{s}} - m_{n\bar{n}} \approx 200-300$
MeV suggests that the primitive $s\bar{s}$ state is in the region
1600-1700 MeV. Hence it is quite possible that the primtive glueball is
in the vicinity of the quarkonium nonet, maybe in the middle of it. Indeed,
the suppression of $K\bar{K}$ in the $f_0(1500)$ decays suggests
a destructive interference between $n\bar{n}$ and $s\bar{s}$
such that $\omega R^2 < 0$. This
arises naturally if the primitive glueball mass is
between those of $n\bar{n}$ and the primitive $s\bar{s}$. As the
mass of $G_0 \rightarrow m_{n\bar{n}}$ or $m_{s\bar{s}}$, the
$K\bar{K}$ remains suppressed though non-zero; thus eventual
quantification of the $K\bar{K}$ signal will be important.

The decay into pairs of glueballs, or states such as $\eta$ that
appear to couple to gluons, is triggered by the perturbation $V_2$.
This can drive decays into $\eta\eta$ and is discussed in ref.\cite{cafe95}.
This breaks the connection between $\eta\eta$ and $K\bar{K}$ that is
a signature for quarkonium as illustrated earlier. The phenomenology
of the $f_0(1500)$ appears to have these features.

 If the $f_0(1550 \pm 50)$ becomes accepted as a
scalar glueball, consistent with the predictions of the lattice, then
searches for the $0^{-+}$ and especially the $2^{++}$ at mass
2.22 $\pm$ 0.13 GeV \cite{teper} may become
seminal for establishing the lattice as a successful calculational laboratory.
There are tantalising indications of a state produced in $\psi \rightarrow
\gamma 0^- 0^-$ at BES whose decays may be consistent with those of
a flavour blind glueball (flavour blind as it is removed from the
prominent quarkonia of the same quantum numbers)\cite{beijing}.

It also adds confidence to the predictions that gluonic degrees of freedom
are excited in the 2GeV mass region when $q\bar{q}$ ``seeds" are already
present. Such states are known as hybrids and these too may be showing up
(see later).

\section{PRODUCTION RATES}

There are two main phenomenological pillars on which glueball
phenomenology now tend to agree. These are their mass spectroscopy
(at least for the lightest few states), and their optimised production
mechanisms. We shall see that the ``interesting" states appear to share
these properties.

Meson spectroscopy has been studied for several decades and the spectrum of
$q\bar{q}$ states has emerged. Why in all this time has it been so hard
to identify glueballs and hybrids if they exist below 2GeV?

Some time ago I suggested \cite{closerev}  this
to be due to the experimental concentration on a restricted class of
production mechanisms and on final states
with charged pions and kaons. We will consider each of these in turn.

Experiments historically have tended to use beams of quarks (contained within
hadrons) hitting targets which are also quark favoured. The emergence
of states made from quarks was thereby emphasised. To enhance any
gluonic signal above the quark ``noise" required one to destroy the
quarks. Hence the focussing on three particular production mechanisms,
\cite{closerev}
in each of which the candidate scalar glueball\cite{cafe95} has been seen.

\begin{enumerate}
\item
Radiative $J/\psi$ decay: $J/\psi  \rightarrow \gamma+G$
\cite{bugg}
\item
Collisions in the central region away from quark beams and target:
$pp \rightarrow p_f(G)p_s$ \cite{Kirk,Gentral}.
\item
Proton-antiproton annihilation where the destruction of quarks
creates opportunity for gluons to be manifested. This is the Crystal
Barrel \cite{Anis}-\cite{Enhan} and E760
\cite{Hasan1,Hasan2} production mechanism in which detailed
decay systematics of $f_0(1500)$ have been studied.
\item
Tantalising further hints come from the claimed sighting \cite{had95}
of the $f_0(1500)$ in decays of the hybrid meson candidate \cite{cp94}
$\pi(1800) \rightarrow \pi f_0(1500) \rightarrow \pi \eta \eta$.
\end{enumerate}

The signals appear to be prominent in decay channels such as
$\eta\eta$ and $\eta\eta'$ that are traditionally regarded as
glueball signatures.  This recent emphasis on
neutral final states (involving $\pi^0$, $\eta$, $\eta'$) was
inspired by the possibiblity that $\eta$ and $\eta'$ are strongly
coupled to glue and reinforced by the earlier concentrations on
charged particles. This dedicated study of neutrals was a new
direction pioneered by the GAMS Collaboration at CERN
announcing new states decaying to $\eta\eta$ and
$\eta\eta'$ \cite{Aldeall}. Note from the decays of quarkonia, fig 4,
the channels $\eta\eta$ and $K\bar{K}$ are strongly correlated for
quarkonia. Thus observation of states that couple strongly to $\eta$
are signatures for non-quarkonia and, to the extent that $\eta$ couples
to glue, may be a glueball signature.

These qualitative remarks are now becoming more quantitative following
work on $\psi$ radiative decays that is currently being extended
\cite{cak,zpli}
  By combining the known B.R.
$(\psi\rightarrow\gamma R$) for any resonance $R$ with perturbative QCD
calculation  of  $\psi\rightarrow\gamma (gg)_R$ where the two gluons are
projected onto the $J^{PC}$ of $R$,
one may estimate the gluon branching
ratio $B(R\rightarrow gg$). One may expect that

\begin{equation}
\begin{array}{lcl}
B(R[Q\bar{Q}] \rightarrow gg)& =& 0(\alpha^2_s) \simeq 0.1\nonumber\\
B(R[G] \rightarrow gg)& =& \frac{1}{2} \; \mbox{to} \; 1\nonumber\\
\end{array}
\end{equation}
Known $Q\bar{Q}$ resonances (such as $f_2$(1270)) satisfy the former; we seek
examples of the latter.

For example, perturbative QCD gives

$$
B(\psi \rightarrow \gamma^3P_J) = \frac{128}{5}  \frac{\alpha \alpha_s}{\pi}
\frac{1}{(\pi^2-9)} \frac{|R^\prime_p (0)  |^2}{m^3 M^2} x|H_J |^2
$$
where $m, M$ are the resonance and $\psi$ masses respectively, $R^\prime_p
(0) $ is the derivative of the $P-$state wavefunction at the origin and the $J$
dependent quantity $x|H |^2$ is plotted in fig 7.  One can manipulate the
above formula into the form, for scalar mesons
$$
10^3  B(\psi \rightarrow \gamma 0^{++}) = (\frac{m}{1.5\; GeV})
(\frac{\Gamma_{R\rightarrow gg}}{96\; MeV})  \frac{x|H|^2}{35}
$$

The analysis of ref. \cite{bugg} suggests
$ B(\psi \rightarrow \gamma f_0 (1500)\simeq 10^{-3})$.
  Thus a very broad $Q\bar{Q}$ state (width
$\sim$ 500 MeV) could be present at this level, but for $f_0$(1500) with
$\Gamma_T$ = 100-150 MeV, one infers $B(f_0\rightarrow gg)$ = 0.6 to 0.9
which is far from  $Q\bar{Q}$.
Such arguments need more
careful study but do add to the interest in the $f_0$(1500).

Thus
the $f_0$(1500) has the right mass and is produced in the right places to be a
glueball and with a strength (in $\psi \rightarrow \gamma f_0$) consistent
with a glueball.
Its total width is out of line with expectations for a $Q\bar{Q}$\cite{cafe95}.
  Its branching
ratios are interesting and may also signify a glueball that is mixed in
with the neighbouring $Q\bar{Q}$ nonet. It is a state for
which data are accumulating and will be worth watching.

\section{THE HYBRID CANDIDATES}

The
origins of the masses of gluonic excitations on the lattice are known only to
the computer.  Those in the flux tube have some heuristic underpinning.  The
$Q\bar{Q}$
are connected by a colour flux with tension 1 GeV/fm which leads to
a linear potential in accord with the conventional  spectroscopy (section 3).

The
simplest glue loop is based on four lattice points that are the corners of a
square.
As lattice spacing tends to zero one has a circle, the diameter is $\simeq$
0.5 fm, the circle of flux length is then $\simeq$ 1.5 fm and, at 1 GeV/fm, the
ballpark
1.5 GeV mass emerges.  In the limit of lattice spacing vanishing, its 3-D
realisation is a sphere, and hence it is natural that this is $0^{++}$.

The next simplest configuration is based on an oblong, one link across and two
links long.
The total length of flux is $\simeq \frac{3}{2} $  larger than the square
and the ensuing mass $\simeq  \frac{3}{2} \times $ 1.5 GeV $\simeq $ 2.2 GeV.
In the
3-D continuum
limit this rotates into a rugby ball shape rather than a sphere.  A
decomposition in spherical harmonics contains $L \geq 0$, in particular
$2^{++}$.
This is by no means rigorous (!) but may help to give a feeling for the
origin of the glueball systematics in this picture, inspired by the lattice.

Finally
one has the prediction that there exist states where the gluonic degrees of
freedom
are excited in the presence of $Q\bar{Q}$.  With the 1 GeV/fm setting the
scale,
one finds \cite{bcs,paton85}
that the lightest of these ``hybrid" states have masses of
order 1 GeV above their conventional $q\bar{q}$ counterparts.  Thus hybrid
charmonium may exist at around 4 GeV, just above the $D\bar{D}$  pair
production
threshold.  More immediately accessible
 are light quark hybrids that are expected
in the 1.5 to 2 GeV range after spin dependent mass splittings are allowed for.

There are
tantalising sightings of an emerging spectroscopy as I shall now review.

It is well known that hybrid mesons can have $J^{PC}$ quantum numbers
 in combinations such as $0^{--},0^{+-},
1^{-+}, 2^{+-}$ etc. which are unavailable to conventional mesons and as
such provide a potentially sharp signature.

It was noted in ref.\cite{kokoski85} and confirmed in ref.\cite{cp95}
that the best opportunity for isolating exotic hybrids appears
to be in the $1^{-+}$ wave where, for the I=1 state with mass around 2 GeV,
partial widths are typically

\begin{equation}
\label{bnlwidth}
 \pi b_1 : \pi f_1 : \pi \rho \;
= \; 170 \; MeV : 60 \; MeV : 10 \; MeV
\end{equation}
The narrow $f_1(1285)$ provides a useful tag for the
$1^{-+} \rightarrow \pi f_1$ and ref.\cite{lee94} has recently reported a
signal
in $\pi^- p \rightarrow (\pi f_1) p$ at around 2.0 GeV
that appears to have a resonant phase.

Note
the prediction  that the $\pi \rho$ channel is
not negligible relative to the signal channel
$\pi f_1$
thereby resolving the puzzle of the production
mechanism that was commented upon in ref. \cite{lee94}.
This state may also have been sighted in photoproduction \cite{utk}
with $M=1750$ and may be the $X(1775)$ of the Data Tables, ref.\cite{PDG}.

A recent development
is the realisation that even when hybrid and conventional mesons
have the {\bf same} $J^{PC}$ quantum numbers, they may still be distinguished
\cite{cp95} due to their different internal structures which give rise to
characteristic selection rules\cite{pene,paton85,cp95}.  As an example
consider the  $\rho(1460)$.

(i) If $q\bar{q}$ in either hybrid or conventional mesons
are in a net spin singlet configuration then the dynamics of the flux-tube
forbids decay into final states consisting only of spin singlet mesons.

For $J^{PC}=1^{--}$  this selection rule
distinguishes between conventional vector mesons
which are $^3S_1$ or $^3D_1$ states and hybrid vector mesons where the
$Q\bar{Q}$ are coupled to a spin singlet. This implies that in the decays of
hybrid $\rho$, the channel
$\pi h_1$ is forbidden whereas $\pi a_1$ is allowed and
that $\pi b_1$ is analogously suppressed for hybrid $\omega$ decays; this
is quite opposite to the case of $^3L_1$ conventional mesons where the
$\pi a_1$ channel is relatively suppressed and $\pi h_1$ or $\pi b_1$
are allowed\cite{busetto,kokoski87}. The extensive analysis of data
in ref.\cite{don2} revealed the clear presence of $\rho(1460)$\cite{PDG}
 with a strong $\pi a_1$ mode but no sign of $\pi h_1$,
in accord with the hybrid situation. Furthermore, ref.\cite{don2}
finds evidence for $\omega (1440)$ with no visible
decays into $\pi b_1$ which again  contrasts with
 conventional $q\bar{q}$ $(^3S_1$ or $^3D_1$) initial states and in accord
with the hybrid configuration.

(ii) The dynamics of the excited flux-tube in the {\bf hybrid} state
suppresses the decay to mesons where the $q\bar{q}$ are $^3S_1$ or $^1S_0$
$``L=0"$ states. The preferred decay channels are to ($L=0$) + ($L=1$)
pairs\cite{paton85,kokoski85}. Thus
in the decays of hybrid $\rho \rightarrow 4\pi$ the $\pi a_1$ content is
predicted to be dominant and the $\rho \rho$ to be absent. The
analysis of ref.\cite{don2} finds such a pattern for $\rho(1460)$.

(iii) The selection rule forbidding ($L=0$) + ($L=0$) final states
no longer operates if the internal structure or size of the two ($L=0$) states
differ\cite{paton85,pene}.
 Thus, for example, decays to $\pi + \rho$, $\pi + \omega$ or $K + K^*$
may be significant in some cases\cite{cp95,cp951},
and it is possible that the {\bf production}
strength could be significant where an exchanged
 $\pi, \rho$ or $\omega$ is involved,
as the
exchanged off mass-shell state may have different structure to the incident
on-shell beam particle.
 This may be particularly pronounced in the case
of {\bf photoproduction} where couplings to $\rho \omega$ or $\rho \pi$
 could be considerable when the $\rho$ is effectively replaced
by a photon and the $\omega$ or $\pi$ is exchanged.
This may explain the production of the candidate exotic
$J^{PC}=1^{-+}$ (ref.\cite{lee94}) and a variety of anomalous
signals in photoproduction.

The first calculation of the widths and branching ratios of hybrid
mesons with conventional quantum numbers is in ref.\cite{cp95}:
the
$0^{-+},2^{-+}$ and the
$1^{--}$ are predicted to be potentially accessible.
It is therefore interesting that each of these  $J^{PC}$ combinations
shows rather clear
signals with features characteristic of hybrid dynamics and which do not
fit naturally into a tidy $Q\bar{Q}$ conventional classification.

We have already mentioned the $1^{--}$.
Turning to the $0^{-+}$ wave,
the VES Collaboration at Protvino confirm their enigmatic
and clear $0^{-+}$ signal in diffractive production with 37 GeV
incident pions on beryllium \cite{had95}. Its mass and decays typify those
expected for a hybrid: $M \approx 1790$ MeV, $\Gamma \approx 200$ MeV
in the $(L=0)$ + $(L=1)$ $\bar{q}q$
 channels $\pi^- + f_0; \; K^- + K^*_0, \; K {( K \pi )}_S $ with no
corresponding strong signal in the kinematically allowed $L=0$ two body
channels $\pi + \rho; \; K + K^*$. This confirms the earlier sighting
by Bellini et al\cite{bellini}, listed in the Particle Data group\cite{PDG}
as $\pi(1770)$.

The resonance also appears to couple as strongly to
the enigmatic $f_0(980)$ as it does to $f_0(1300)$,
which was commented upon with some surprise in ref. \cite{had95}.
This may be natural for a hybrid at this mass due to the
predicted dominant $KK_0^*$ channel which will feed
the $(KK\pi)_S$ (as observed \cite{had95}) and hence the channel
$\pi f_0(980)$ through the strong affinity of $K\bar{K} \rightarrow f_0(980)$.
Thus the overall expectations for hybrid $0^{-+}$ are in line with
the data of ref.\cite{had95}. Important tests are now that there should be
a measureable decay to the $\pi \rho$ channel with only a small
$\pi f_2$ or $KK^*$ branching ratio.
At the Hadron95 conference it was learned
that in the $\pi \eta\eta$ final state the glueball
candidate is seen: $\pi(1.8) \rightarrow \pi f_0(1500)
\rightarrow \pi \eta\eta$.

This leaves us with the $2^{-+}$.
There are clear signals of unexplained activity in the
$2^{-+}$ wave in several experiments for which a hybrid interpretation
may offer advantages. These are discussed in ref.\cite{cp95}.

These various signals in the desired channels provide a potentially consistent
picture. The challenge now is to test it. Dedicated high statistics
experiments with the power of modern detection and analysis should re-
examine
these channels.  Ref.\cite{cp951} suggests that the hybrid couplings are
especially
favourable in
{\it low-energy} photoproduction and as such offer a rich opportunity
for the programme at an upgraded CEBAF
 or possibly even at HERA. If the results of ref.\cite{atkinson}
are a guide, then photoproduction may be an important gateway at a range of
energies and the channel $\gamma + N \rightarrow (b_1 \pi) + N$ can
discriminate
hybrid $1^{--}$ and $2^{-+}$ from their conventional counterparts.

Thus to summarise, we suggest that data are consistent with the existence of
low lying multiplets of hybrid mesons based on the mass spectroscopic
predictions of ref.\cite{paton85} and the production and decay dynamics of ref.
\cite{cp95}. Specifically the data include

\begin{eqnarray}
 0^{-+} & (1790 \; MeV; \Gamma = 200 \; MeV) & \rightarrow
\hspace{0.2cm} \pi f_0 ; K\bar{K}\pi \\
\nonumber
 1^{-+} & (\sim 2 \; GeV; \Gamma \sim 300 \;  MeV)  & \rightarrow
\hspace{0.2cm}  \pi f_1 ; \pi b_1 (?) \\
\nonumber
 2^{-+} & (\sim 1.8 \; GeV; \Gamma \sim  200 \; MeV)  & \rightarrow
\hspace{0.2cm}  \pi b_1; \pi f_2 \\
\nonumber
 1^{--} & (1460 \; MeV; \Gamma \sim 300 \; MeV)  & \rightarrow
\hspace{0.2cm}  \pi a_1
\end{eqnarray}

Detailed studies of these and other relevant channels are called for together
with analogous searches for their hybrid charmonium analogues, especially in
photoproduction or $e^+ e^-$ annihilation.

\section{RADIALOGY}
If these states are not glueballs and hybrids, what are they? On masses alone
they could be radial excitations as we have already noted. The decay patterns
 have been seen to fit well with gluonic excitations but we need to close
the argument by considering the decays in the radial hypothesis. Only if
the hybrid succeeds and radial fails can one be sure to have a convincing
argument.

As an illustration consider the $0^{-+}$ (1800) which could be either
the hybrid or the radial
$3S (^1S_0)$ quarkonium (denoted $\pi_{RR}$).  In fig 8 we see the width in an
S.H.O. calculation as a function of the oscillator strength $\beta$.
The $\pi \rho$ channel is small near the preferred value of $\beta \simeq$ 0.35
- 0.4 GeV and so both radial and hybrid share the property of suppressed $\pi
\rho$  and, to some extent, $KK^*$.  It is therefore encouraging that data
 show a
clear absence of $\pi \rho$  in the 1800 region in
contrast to $\pi f_0$  which
shows two clear bumps at 1300 and 1800 MeV.  Notice that for $\pi_{RR}$ the
$\pi f_0$ is small for all $\beta$ in dramatic contrast to the hybrid, where
this
channel is predicted to dominate, and also apparently in contrast to data.
The
$\pi \rho_R$ (radial $\rho$) is predicted to be large for $\pi_{RR}$  and hence
one would expect a significant branching ratio $\pi_{RR}\rightarrow  \pi
\rho_R \rightarrow 5 \pi$  which is not apparent in data though more study is
warranted.

A discriminator between   $\pi_{RR}$ and hybrid $\pi_H$ is the
$\rho \omega$  channel.  This is a dominant channel for  $\pi_{RR}$ whereas
it is predicted to be absent for $\pi_H$ \cite{cp95}.

Another example that distinguishes hybrid and radial is in the $1^{++}$ sector.
There is a clear signal in $\pi f_1$ \cite{lee94}; $a_{1H}$ is forbidden to
decay into
$\pi b_1$ due to the spin selection rule \cite{cp95} whereas $a_{1R}\rightarrow
\pi b_1$ with a branching ratio equal to $a_{1R} \rightarrow \pi f_2$ over the
full $\beta$ range and moreover $\Gamma (a_{1R} \rightarrow \pi b_1)
\gapproxeq 2 \Gamma (a_{1R} \rightarrow \pi f_1)$.  (see fig 9)  The $\pi f_2$
channel may be easier experimentally.  In any event we see that there are
characteristic differences in the branching ratios for radials and hybrids
states that
should enable a clear separation to be made.

After years of searching, at last we have some potential candidates for mesons
where the gluonic degrees of freedom are excited. Furthermore there are
some clear selection rules
and other discriminators in their decay branching ratios that can help to
verify
their existence and thereby complete the strong QCD sector of the
standard model.

\vskip 0.2in
\noindent {\bf Acknowledgements}

We thank the organisers for having arranged such a stimulating school, for
having made me feel at home by arranging British weather
and for  introducing us to the delightful beaches of Ubatuba.

\pagebreak

\pagebreak
\section*{Figure Captions}

\begin{figure}[h]
\vspace{1mm}
\caption[]{Comparative spectroscopies of Coulomb, linear and oscillator
potentials}
\end{figure}

\begin{figure}[h]
\vspace{1mm}
\caption[]{Template for $q\bar{q}$ spectroscopy; (a) bottomium, (b) charmonium
(c) u and d flavours}
\end{figure}

\begin{figure}[h]
\vspace{1mm}
\caption[]{The lightest L=0-3 $q\bar{q}$ ($q = u,d$) and $_{\Lambda} L = _1P$
hybrid masses from Monte Carlo (after ref.\cite{bcs}). Square brackets denote
masses used as input}
\end{figure}

\begin{figure}[h]
\vspace{1mm}
\caption[]{$\gamma_{ij}^2$ as a function of $\alpha$ for $\rho=1$
(a)
and $\rho=0.75$ or $1.25$ (b) for quarkonium decay (up to a common
multiplicative factor). Dotted line:
$\pi\pi$; dash-dotted line: $\KKbar$; dashed line:
$\eta\eta'$; solid line: $\eta\eta$.}
\label{alpha}
\end{figure}

\begin{figure}[h]
\vspace{1mm}
\caption[]{Contributions to gluonium decay: $QQ\bar{QQ}$
(a), $GG$ (b), $\qqbar$ (c) and interpretation as $Q\bar{Q}$ mixing
(d)
involving the energy denominator $E_G-E_{Q\bar{Q}}$}
\label{3graphs}
\end{figure}

\begin{figure}[h]
\vspace{1mm}
\caption[]{Glueballs, quarkonia and perturbations:
(a) primitive $Q\bar{Q}$ and (b) primitive glueball $G_0$ in flux tube
simulation of lattice QCD; perturbation $V_1$ (c) and $V_2$ (d);
the effect of $V_1$ on $Q\bar{Q}$ is shown in (e), and on $G$ is shown in
(f); the effect of $V_2$ on $G$ is shown in (g).}
\label{Rivsx}
\end{figure}

\begin{figure}[h]
\vspace{1mm}
\caption[]{$x|H|^2$ versus $x \equiv 1-\frac{m^2}{M_\psi^2}$. Solid line is
$0^{++}$, dashed line is $2^{++}$}
\end{figure}

\begin{figure}[h]
\vspace{1mm}
\caption[]{Partial widths of a $\pi_{RR}(1800)$ second radial excitation;
$^3P_0$ model normalised to $f_2(1270)$ width with wavefunction parameter
$\beta$ variable}
\end{figure}

\begin{figure}[h]
\vspace{1mm}
\caption[]{As previous figure but for $a_{1R}(1700)$ radial excitation}
\end{figure}

\end{document}